\documentclass[preprint,showpacs,preprintnumbers,amsmath,amssymb,11pt]{revtex4}
\usepackage{graphicx}
\usepackage{dcolumn}
\usepackage{bm}
\usepackage{color}
\newcommand{\be}{\begin{equation}}
\newcommand{\en}{\end{equation}}
\newcommand{\bea}{\begin{eqnarray}}
\newcommand{\ena}{\end{eqnarray}}
\begin{document}
\title{Single-field inflation {\emph {\lowercase { \`{a} la}}} generalized Chaplygin gas}
\author{Sergio del Campo}
\email{sdelcamp@ucv.cl}
\affiliation{ Instituto de F\'{\i}sica, Pontificia Universidad Cat\'{o}lica de Valpara\'{\i}so, Casilla 4059, Valpara\'{\i}so, Chile.}
\date{\today}
\begin{abstract}
In the simplest scenario for inflation, i. e. in the single-field inflation,
it is presented an inflaton
field with properties equivalent to a generalized Chaplygin
gas. Their study is performed using the Hamilton-Jacobi approach
to cosmology. The main results are contrasted with the
measurements recently released by the Planck data, combined with
the WMAP large-angle polarization. If the measurements released by
Planck for the scalar spectral index together with its
running are taken into account it is found a value for the $\alpha$-parameter
associated to the generalized Chaplygin gas given by $\alpha =
0.2578 \pm 0.0009$.
\end{abstract}
\pacs{98.80.Cq}
\maketitle

\section{Introduction}

The inflationary paradigm has been proposed as a good approach for
solving most of the cosmological "puzzles" \citep{linde90a}. Apart
of solving these cosmological problems, inflation produces the
seeds that, in the course of the subsequent eras of radiation and
matter dominance, developed into the cosmic structures (galaxies
and clusters thereof) that we observe today. In fact, the present
popularity of the inflationary scenario is entirely due to its
ability to generate a spectrum of density perturbations which lead
to structure formation in the universe. In essence, the conclusion
that all the observations of microwave background anisotropy
performed so far support inflation, rests on the consistency of
the anisotropy with an almost Harrison-Zel'dovich power spectrum
predicted by most of the inflationary universe scenarios
\citep{WMAP03} and corroborated by the measurements recently
released by the Planck combined with the WMAP large-angle
polarization\cite{Planck2013}.

The essential feature of any inflationary universe model proposed
so far is the rapid (accelerated) but finite period of expansion
that the universe underwent at very early times in its evolution.
The implementation of the inflationary universe model rests on the introduction of a scalar
inflaton field, $\phi$. The evolution of this field becomes
governed by its scalar potential, $V(\phi)$, via the Klein-Gordon
equation. Thus, this equation of motion, together with the
Friedmann equation, obtained from Einstein general relativity
theory, form the most simple set of field equations, which could
be applied to obtain inflationary solutions. But, in order to do
this it is necessary to give an explicit expression for the scalar
inflaton potential $V(\phi)$. However, in simple cases result very
complicated to find solutions, even in the situation in which it
is applied the so-called slow-roll approximation, where the
kinetics terms is much smaller than the potential energy, i.e.
$\dot{\phi}^2 \ll V(\phi)$, together with the approximation $\mid
\ddot{\phi} \mid \ll H \mid \dot{\phi} \mid$. This approach is
usually refereed as the {\em single-field inflation},
whose definition encompasses the type of models where the inflationary phase becomes described
by a single inflaton scalar field.

One way of finding inflationary solutions out of the slow-roll
approximation is giving the functional form of the Hubble
parameter in term of the inflaton field, i.e. $H(\phi)$, the so
called generating fuction\cite{L91}. This approach presents some
advantages when compared with the slow-roll approximation: first
of all, the form of the potential is deduced, and secondly, since an
exact solution is obtained, then, application to the final period
of inflation is possible. We should note that in this period the
kinetic term associated to the inflaton
field in the Friedmann equation becomes important. Certainly, this
approach becomes necessary  when studying the final stage of
inflation, i.e. during the reheating phase. This method, that we will consider
here, is usually referred as the Hamilton-Jacobi scheme to
cosmology\cite{GS88,LetAl97,dC12}.

In this paper we would like to study the consequences of
considering an inflationary universe model in which the inflaton
field presents characteristic of a generalized Chaplygin
gas\cite{B01,B02,G03,B04}, with Equation of state \be p_{\phi} = -
B \rho_{\phi}^{-\alpha}, \label{EoSCh} \en where  $B$ is a positive constant and $\alpha$
a parameter that we call the generalizes Chaplygin gas parameter, which will be considered
to lie in the range $0< \alpha < 1$, and $p_\phi$ and $\rho_\phi$ are defined,
as usual, by expressions $p_\phi \equiv\frac{1}{2}\,\dot{\phi}^2 -
V(\phi)$ and $\rho_\phi \equiv \frac{1}{2}\,\dot{\phi}^2 + V(\phi)$,
respectively. The usual Chaplygin gas is obtained when the
parameter $\alpha$ becomes equal to one. Because the range that we have
set for this parameter does not provide equality we will not include the possibility
of considering the usual Chaplygin gas in our analysis. Moreover, it is possible to extend our analysis to
include values of the parameter $\alpha$ which lie outside the range specified above,
even by taking negative values, but our results do not change much. For simplicity, we shall restrict
ourselves to considerer this parameter to lie in the range that
we have specified above.

The sort of matter related to the usual Chaplygin gas, and its
generalizations, have been extensively studied, and considered as
one of the main matter components of the universe at present
time\cite{KMP}. It has been established that this kind of
component could be considered as the unification of dark matter
and dark energy\cite{B01}. On the other hand, the relation between
a perfect fluid, which can be related to a Chaplygin gas together
with the corresponding scalar field models were considered in
Refs\cite{PMT04,GKMPS05,U13}. The different parameters that enter
into the model have been confronted with the recent observational
data coming out of gravitational lensing, the baryon acoustic
oscillation, the Cosmic Microwave Background Radiation from the
Wilkinson Microwave Anisotropy Probe (WMAP7) result\cite{L11},
Gamma-ray bursts\cite{F11}, power spectrum observational
data\cite{Fa11}, the Constitution data set of type supernovae Ia
cosmic probes, together with sample of cosmology-independent long
gamma-ray bursts calibrated using their Type I Fundamental Plane,
as well as the Union 2.1 set and observational Hubble parameter
data\cite{L12}, the combination of Chandra observations of the
x-ray luminosity of galaxy clusters, together with independent
measurements of the baryonic matter density, the latest
measurements of the Hubble parameter as given by the HST Key
Project, and data of the Supernova Cosmology Project\cite{C04}.
Also, it was applied the the statefinder parameters\cite{S03}.
Furthermore, combinations of some of these measurements have been
made as well\cite{W13}. On the other hand, the generalized
Chaplygin gas has been studied by considering non-adiabatic
pressure perturbations, where it was found that there are no
instabilities which have been found in previous adiabatic
Chaplygin-gas models\cite{B13}. Also, a successful description was
obtained for generalized Chaplygin gas in thawing
models\cite{dC13}.

One of the interesting features of the generalize Chaplygin gas is
the connection that it has with string theory. Effectively, it can
be obtained from the Nambu-Goto action for a D-brane moving in a
(D+1,1)-dimensional spacetime in the light cone
parametrization\cite{varios01}. This open a window for applying
this sort of theory at early time in the evolution of the
universe. The idea of taking an early phase of the universe
in which the main component corresponds to a Chaplygin gas has
been considered in Ref. \cite{varios001}. The tachyon-Chaplygin
inflationary universe model was studied in Ref. \cite{dC08}
under the slow-roll approximation.

The out line of this paper is as follows. We first describe
exact solutions to inflationary universe
models within the framework of the so-called Hamilton-Jacobi approach to
cosmology. This will be carried out in section II. Then, in section III
we apply this approach to the generalized Chaplygin
gas inflationary universe model. In section IV we study the
corresponding scalar and tensor perturbations. Finally, we
conclude our work in section V.



\section{The Hamilton-Jacobi approach to inflation}
The simplest model that describes inflationary universes in
general relativity presents two main ingredients: the flat FRW
metric and the single inflaton scalar field, $\phi$.

In the scheme of Hamilton-Jacobi approach the fundamental
Equations of the model are described by the Klein-Gordon Equation,
which satisfies the inflaton scalar field and it given by \be \ddot{\phi} + 3 H
\dot{\phi} + V'(\phi) = 0, \label{KG01} \en where  $V'$ represents
a derivative with respect to $\phi$, and the Friedmann Equation
given by
\be H^2 = \frac{8 \pi}{3\,m_{Pl}^2} \left[\frac{1}{2}\dot{\phi}^2
+ V(\phi)\right] \label{f}.\en
Here, $m_{Pl}^2 = 1/G$ is the Planck mass.

It is not hard to see from Eqs. (\ref{KG01}) and (\ref{f}) that the following
relationship is satisfied
\be \dot{\phi} = -\,\left(\frac{m_{Pl}^2}{4 \pi}\right) \,H',
\label{DotPhi}\en where $H'$ represents a derivative with respect to
$\phi$. Since it is assumed that the generating function $H(\phi)$
it is known, we can add that this latter relation allows us to
obtain an explicit expression for the inflaton field in terms of
the cosmological time $t$ in cases where it is possible to reverse
the expression of the scalar field, $\phi$.

Also, it is not hard to see that it is found that
\be a\,H =
-\left(\frac{m_{Pl}^2}{4\,\pi}\right)\,a'\,H',\label{15}\en from
which we get that \be a(\phi) =
\exp\left\{-\frac{4\,\pi}{m_{Pl}^2}\,\int\frac{H}{H'}\,d\phi
\right\},\label{16} \en
i.e. the scale factor $a$ as a function of the inflaton field $\phi$, and thus,
assuming that we have the scalar field as a function of time, then we can obtain
the scale factor as a function of the cosmological time.

The scalar potential $V(\phi)$ becomes given by \be V(\phi)=
\left(\frac{3 \, m_{Pl}^2}{8\,\pi}\right)\left[H^2 -
\frac{3\,m_{Pl}^2}{12\,\pi}\left(H'\right)^2\right]. \label{VCh}
\en
We should note that in the slow-roll approximation this potential becomes
$\displaystyle V(\phi) \simeq
\left(\frac{3 \, m_{Pl}^2}{8\,\pi}\right)\,H^2$

On the other hand, we see that the acceleration Equation for the
scale factor results to be
\be \frac{\ddot{a}}{a}=
H^2\left[1-\epsilon_{_{H}}\right],\label{dda} \en
where the function $\epsilon_{_{H}}$ corresponds to
\be \epsilon_{_{H}} \equiv - \frac{d \ln {H}}{d \ln {a}} =
\left(\frac{m_{_{Pl}}^2}{4
\,\pi}\right)\,\left(\frac{H'}{H}\right)^2.\label{e} \en
From this latter expression we can see that this definition,
called the {\it first Hubble hierarchy parameter}, gives
information about the acceleration of the universe. During
inflation this parameter satisfies the bound
$\epsilon_{_{H}} <1$, and inflation ends
when $\epsilon_{_{H}}$ takes the value equal to one. In the next
section we will use this parameter for describing scalar and
tensor perturbations.

One interesting quantity in characterizing inflationary universe
models is the amount of inflation. Usually, this quantity is
defined by
\be N(t) \equiv \ln
{\frac{a\left(t_{end}\right)}{a(t)}},\label{NCh} \en
where $a\left(t_{end}\right)$ corresponds to the scale factor
evaluated at the end of inflation. By considering expressions
(\ref{DotPhi}) and (\ref{e}) we write \be N(\phi) \equiv
\int_t^{t_{end}} {H \,dt} =
\left(\frac{4\,\pi}{m_{_{Pl}}^2}\right)\int^{\phi}_{\phi_{end}}
\frac{H}{H'}\,d\phi = \int^{\phi}_{\phi_{end}}
\frac{1}{\epsilon_{_H}}\,\frac{H'}{H}\,d\phi. \label{N2} \en
Here, $\phi_{end}$ represent the value of the scalar field at the
end of inflation. Its value is determined by imposing that
$\epsilon_{_H}\left(\phi_{end}\right) = 1$. We interpret the parameter
$N$ as the number of e-folding before inflation ends.

It seems to be more appropriated to describe the amount of
inflation in terms of the comoving Hubble length, $1/(a H)$ than
in terms of the scale factor only. In this case the amount of
inflation becomes\cite{LPB94}
\be \overline{\text{N}} = \ln
{\frac{a\left(t_{end}\right)\,H\left(t_{end}\right)}{a(t)\,H(t)}},
\label{N3} \en
which results into
\be \overline{\text{N}}(\phi) = \int^{\phi}_{\phi_{end}}
\left(\frac{1}{\epsilon_{_{H}}}-1\right)\,\frac{H'}{H}\,d\phi.\label{N4}
\en
Note that, in general, $\overline{\text{N}}(\phi)$ is smaller that
$N(\phi)$ and only in the slow-roll limit they coincide. In the following, and for simplicity,
we will restrict ourselves to consider the amount of inflation given by $N(\phi)$.

In the description of inflation it is convenient to show that
their solutions are independent from their initial conditions.
This ensure the true predictive power that presents any
inflationary universe model, otherwise the corresponding physical
quantities associated with the inflationary phase, such that the
scalar or tensor spectra, would depend on these initial
conditions. Thus, with the purpose of being predictive, any
inflationary model needs that their solutions present an attractor
behavior, in the sense that solutions with different initial
conditions should tend to a unique solution\cite{SB90}.

Let us start by considering a linear perturbation, $\delta
H(\phi)$, around a given inflationary solution, expressed by
$H_0(\phi)$. In the following we will refer to this quantity as
the background solution, and any quantity with the subscript zero
is assumed to be evaluated taking into account the background
solution. Therefore, at first order on $\delta H(\phi)$, we get
from the field Equations (\ref{KG01}) and (\ref{f}) that
\be \delta H \simeq \left.\frac{1}{3}\left(\frac{m_{_{Pl}}^2}{4
\pi}\right)\frac{H'}{H}\right|_0\delta H'\label{p1},\en
This latter expression can be solved for getting
\be \delta H(\phi) = \delta H (\phi_{_i})
\exp\int_{\phi_{_i}}^\phi \left.
\left(\frac{3}{\epsilon_{_{H}}}\right) \frac{H'}{H}\right|_0
d\phi,\label{ps} \en
where $\phi_{_i}$ corresponds to some arbitrary initial value of
$\phi$.  Since $d \phi$ and $H'$ have opposite signs (assuming
that $\dot{\phi}$ does not change sign due to the perturbation
$\delta H$) the linear perturbations tend to vanish
quickly\cite{LPB94}.

In the following we want to apply this scheme to the case in which
the inflaton field corresponds to a generalized Chaplygin gas
fluid whose equation of state is governed by expression
(\ref{EoSCh}).


\section{Inflation {\emph {\lowercase { \`{a} la}}} Chaplygin}

In this section we describe the inflationary model by using a
fluid which presents the properties of a generalized Chaplygin gas, where the
equation of state of this fluid corresponds to that specified by
equation (\ref{EoSCh}). In order to do this we start by taking
that the generating function, $H(\phi)$ is given by \be
H(\phi) =H_0 \cosh^{\frac{1}{1+\alpha}}\left[\sqrt{\frac{6
\pi}{m_{Pl}^2}}(1+\alpha)(\phi - \phi_0)\right], \label{HCh} \en
where $H_0$, the value of $H$ when $\phi=\phi_0$, is a constant
given by $\displaystyle H_0 = \sqrt{\frac{8
\pi}{3\,m_{Pl}^2}}\,B^{\frac{1}{2(1+\alpha)}}$, with $B$ and $\alpha$
introduced previously in Eq. (\ref{EoSCh}). For simplicity from
now on we will use the following dimensionless variable as the
inflaton scalar field: $\Phi(\phi) \equiv \sqrt{\frac{6
\pi}{m_{Pl}^2}}(\phi - \phi_0)$.

In order to see that the expression above corresponds to a fluid
which has an equation of state related to a Chaplygin gas, we take
into account that the pressure and the energy density are given by
\be p_{\phi}= \frac{3 m_{Pl}^2}{8 \pi}\,H^2\left[\frac{m_{Pl}^2}{6
\pi}\,\left(\frac{H'}{H}\right)^2-1\right] \label{Pch} \en and \be
\rho_{\phi} = \frac{3 m_{Pl}^2}{8 \pi}\,H^2,\label{RhoCh} \en
respectively. Now, by using Eq. (\ref{HCh}) into these two latter
Equations we obtain that \be p_{\phi} = - \frac{3 m_{Pl}^2}{8
\pi}\,H_0^2\,\cosh^{-\frac{2 \alpha}{1+
\alpha}}\left[(1+\alpha)\Phi\right]\label{pCh2} \en and \be
\rho_{\phi} = H_0^2
\cosh^{\frac{2}{1+\alpha}}\left[(1+\alpha)\Phi\right],
\label{RhoCh2} \en respectively. From Eq. (\ref{RhoCh2}) we get
$\displaystyle \cosh\left[(1+\alpha)\Phi\right]$ as a function of
$\rho_{\phi}$ and then substituting into Eq. (\ref{pCh2}) we
obtain the Equation of state related to the generalized Chaplygin
gas, as expressed by Eq. (\ref{EoSCh}). In this way, the
generating function expressed by Eq. (\ref{HCh}) corresponds
effectively to a generalized Chaplygin gas generating function.

From Eq. (\ref{DotPhi}) together with Eq. (\ref{HCh}) we get that

\be \csc^{\frac{1}{1+\alpha}}\left[(1+\alpha)\Phi\right] {}_2
F_1\left[\frac{1}{2(1+\alpha)},\frac{1}{2(1+\alpha)};\frac{3+2\alpha}{2(1+\alpha)};
-\csc^2\left[(1+\alpha)\Phi\right]\right] =-\sqrt{\frac{3
m_{Pl}^2}{8 \pi}}\,H_0 t, \label{Phit} \en
where ${}_2 F_1\left[a,b;c;z\right]$ represents the hypergeometric
function. This latter expression gives information on how the
inflaton field evolves with the cosmological time. In the same way
we get that the scale factor, $a(t)$, results to be given by
\be a(\phi) = a_i \left(\frac{\sinh\left[(1+\alpha)\Phi\right]}{
\sinh\left[(1+\alpha)\Phi_i\right]}\right)^{\frac{2}{3}},
\label{aCh} \en
where $a_i$ is the value of the scale factor when the inflaton
field has the value $\phi_i$, and $\Phi_i =\Phi(\phi_i)$.

\begin{figure}[th]
\centering
\includegraphics[width=12cm,angle=0,clip=true]{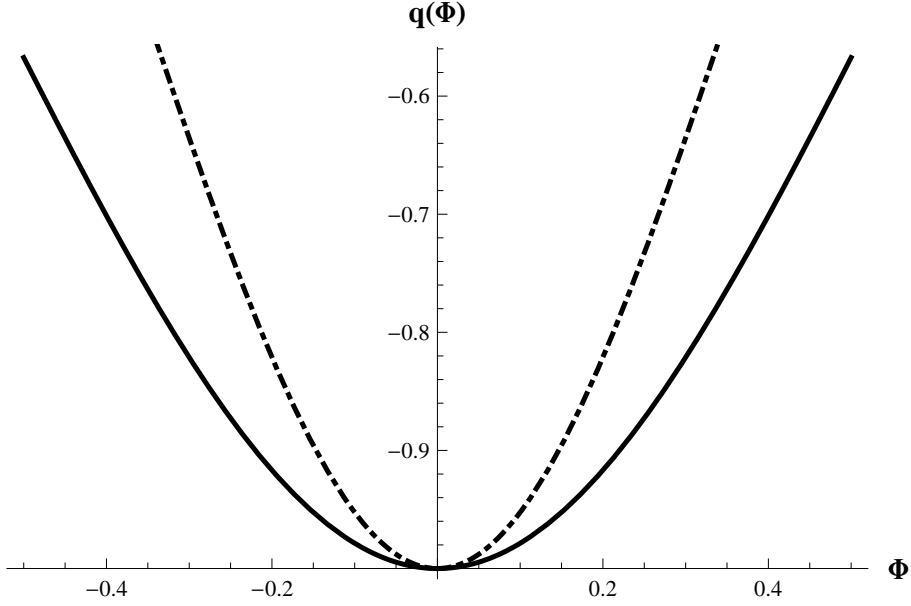}
\caption{The deceleration parameter $q$ as a function of the
dimensionless scalar field $\Phi \equiv \sqrt{\frac{6
\pi}{m^2_{Pl}}}\,(\phi - \phi_0)$. Here we have taken the values
for the generalized Chaplygin gas parameter $\alpha= 0.2$
(dot-dashed line) and $\alpha= 0.8$ (solid line).} \label{fig1}
\end{figure}

In order to see that we have an inflationary period,  we introduce
the deceleration parameter $q$, defined as $\displaystyle q=
-\frac{\ddot{a}\,a}{\dot{a}^2}$, which in terms of the Hubble
parameter it becomes $ \displaystyle q = \left(\frac{m_{Pl}^2}{4
\pi}\right)\,\left(\frac{H'}{H}\right)^2 - 1$, which gives \be q =
-1 + \frac{3}{2}\,\tanh^2\left[(1+\alpha)\Phi\right]. \label{qCh}
\en We note that this quantity becomes negative for the values
$\displaystyle \phi \leq \phi_0\, \pm \,
\frac{1}{1+\alpha}\sqrt{\frac{1}{6\,\pi}}\,
{\text{arctanh}}\left(\sqrt{\frac{2}{3}}\right)\,m_{Pl}$. The
equal sign corresponds to when the $q$ parameter vanishes, i.e.
when inflation ends. Figure \ref{fig1} shows how the parameter $q$
changes as a function of the dimensionless field $\Phi$. There, we
have taken two different values of the parameter $\alpha$. The
dot-dashed line corresponds to the value $\alpha = 0.2$ and the
continuous line to $\alpha = 0.8$. These curves show that the
universe is accelerating, since the parameter $q$ turns out to be
negative, at least for the values that we have considered here.

From expressions (\ref{VCh}) together with Eq. (\ref{HCh})  we
obtain for the scalar potential
\be
 V(\phi) = \,V_0 \cosh^{-\frac{2\,\alpha}{1+\alpha}}\,
 \left[(1+\alpha)\Phi\right]
 \left(1+ \cosh^2\left[(1+\alpha)\Phi\right]\right),
\label{V2Ch}
\en
where $V_0 = \frac{1}{2}B^{\frac{1}{1+\alpha}}$. In the slow-roll
approximation, i.e. when $\dot{\phi}^2 \ll V(\phi)$ together with
$\mid\ddot{\phi}\mid \ll \mid dV(\phi)/d\phi \mid$, the scalar
potential reduces to
\be V_{_{s-r}}(\phi) \simeq V_0 \cosh^{\frac{2}{1+\alpha}}\,
 \left[(1+\alpha)\Phi\right]. \label{VrsCh} \en
\begin{figure}[th]
\centering
\includegraphics[width=12cm,angle=0,clip=true]{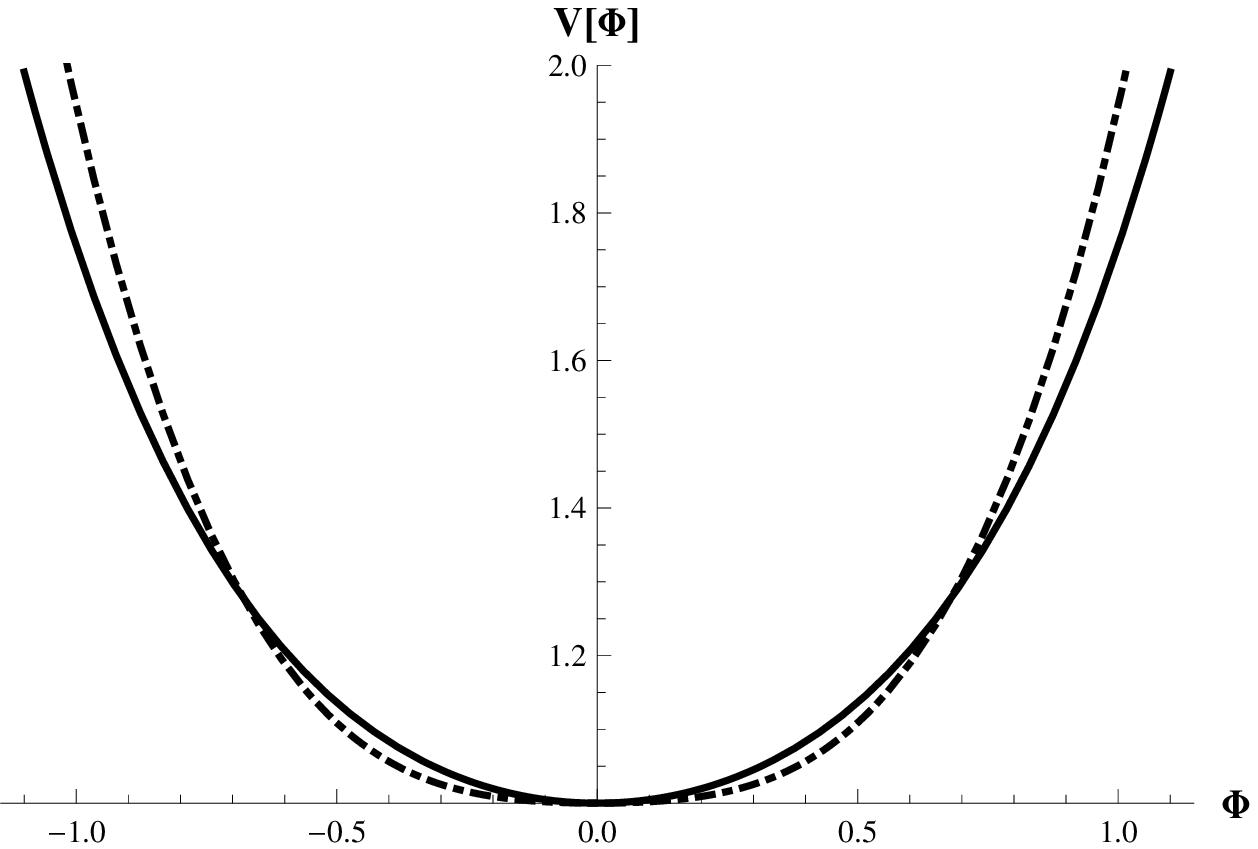}
\caption{Plots of the scalar potentials, $V(\Phi)$, as a function
of the dimensionless scalar field, $\Phi \equiv \sqrt{\frac{6
\pi}{m^2_{Pl}}}\,(\phi - \phi_0)$. The solid line represents the
scalar potential, expressed by Eq. (\ref{V2Ch}) for the value
$\alpha=0.2$. The dot-dashed line represents the same potential,
but for the value $\alpha = 0.8$. The scalar potential  $V(\Phi)$
is expressed as a multiple of the constant $\displaystyle
V_0\equiv \frac{1}{2}B^{\frac{1}{1+\alpha}}$.} \label{fig2}
\end{figure}
Figure \ref{fig2} depicts the shape of the potential for the exact
case, expressed by Eq. (\ref{V2Ch}) as a function of the
dimensionless scalar field $\Phi$. Here, we have plotted the
potential for two values of the parameter $\alpha$. The dot-dashed
and the solid lines correspond to $\alpha = 0.8$ and $\alpha =
0.2$, respectively.

One interesting quantity in characterizing inflationary universe
models is the amount of inflation. This quantity was introduced
above (see Eq. (\ref{NCh})) and together with expression
(\ref{aCh}) we obtain that
\be N(\phi) =\frac{2}{3}\,\ln\left\{\frac{\sinh\left[(1+\alpha)
\Phi_e\right]}{\sinh\left[(1+\alpha)\Phi\right]}\right\},
\label{N2Ch} \en where $\Phi_e=\Phi(\phi_e)$, with $\phi_e$
representing the value of the inflaton field at the end of
inflation. This value is obtained by demanding that the {\em first
Hubble hierarchy parameter} $\epsilon_{_{H}}$ becomes equal to one
at the end of inflation. This parameter becomes determined by
considering its definition, Eq. (\ref{e}), together with
expression (\ref{HCh})
\be \epsilon_{_{H}}(\phi) = \frac{3}{2} \,
\tanh^2\,\left[(1+\alpha)\Phi\right].\label{eCh} \en Now,
requiring that $\epsilon_{_{H}}(\phi_e) = 1$ we get that for the
dimensionless scalar field $\displaystyle \Phi_e =
\frac{1}{1+\alpha}\text{arctanh}\left(\sqrt{\frac{2}{3}}\right)$,
or equivalently, for the inflaton field, $\phi_e = \phi_0\, \pm \,
\frac{1}{1+\alpha}\sqrt{\frac{1}{6\,\pi}}\,{\text{arctanh}}
\left(\sqrt{\frac{2}{3}}\right)\,m_{Pl}$. Note that, as specified
above, this value coincides with that obtained when the parameter
$q$ gets vanished, i.e. when inflation ends.

To study the attractor behavior of the model we consider a
linear perturbation, $\delta H(\phi)$, around a given solution,
just as was specified above. Thus, from Equations (\ref{ps}),
(\ref{HCh}) and (\ref{eCh}) we get that
\be \delta H(\phi)
=\left(\frac{\sinh\left[(1+\alpha)\Phi_i\right]}{
\sinh\left[(1+\alpha)\Phi\right]}\right)^{\frac{2}{1+\alpha}}
\delta H (\phi_{_i}). \label{deltaCh} \en Here, it is assumed that
$\Phi \geq \Phi_i$ is satisfied, and thus, the ratio
$\displaystyle \frac{\sinh\left[(1+\alpha)\Phi_i\right]}{
\sinh\left[(1+\alpha)\Phi\right]}$ decreases rapidly, as
the scalar field $\Phi$ increases. In this way, the linear perturbation
tends to vanish rapidly. This can be seen more clearly, since the
factor accompanying to $\delta H (\phi_{_i})$ is nothing but
$(a/a_i)^{-\frac{3}{1+\alpha}}$ (see expression (\ref{aCh})),
which during inflation the quantity in this bracket increases at
least 70 e-fold, leaving $\delta H(\phi)$ very small.
\section{Scalar and tensor perturbations}

Quantum fluctuations are amplified during inflation. They are
stretched to astrophysical scales by the rapid expansion that
universe presents at early time. In general terms any model of
inflation generate two types of perturbations, density
perturbations (which come from quantum fluctuations in the scalar
field, together with the corresponding scalar metric
perturbation\cite{MCh81,L80}), and relic gravitational waves which
are tensor metric fluctuations\cite{G75}. The former experience
gravitational instability and lead to structure
formation\cite{MFB92}, while the latter predicts a stochastic
background of relic gravitational waves which could influence the
cosmic microwave background anisotropy via the presence of
polarization in it\cite{GW}.

In order to describe these perturbations we need to introduce a
series of parameters which are known as the {\em Hubble hierarchy
parameters}. We have already defined one of them, the so-called
{\em first Hubble hierarchy parameter}, which is expressed by
expression (\ref{eCh}). The {\em second Hubble hierarchy
parameter}, $\eta_{_H}$, is defined by
\be \eta_{_H} \equiv -\frac{d\,\ln H'}{d\,\ln a}=
\frac{m_{Pl}^2}{4\,\pi}\,\left(\frac{H''}{H}\right), \label{etaCh}
\en
from which we get that for the generalized Chaplygin gas it
becomes
\be \eta_{_H} = \frac{3}{4}\,\left(1+2\alpha +
\cosh\left[2\,(1+\alpha)\Phi\right]\right)\,{\text{
sech}}^2\left[(1+\alpha)\Phi\right]. \label{eta2Ch} \en

 The {\em third Hubble hierarchy parameter}, $\xi_{_{H}}^2$
is defined by
\be \xi_{_{H}}^2 \equiv
\left(\frac{m_{_{Pl}}^2}{4\,\pi}\right)^2\,
\left(\frac{H'''\,H'}{H^2}\right),\label{xiCh} \en
which results to be
\begin{align}\xi_{_{H}}^2 = \frac{9}{8}\,\left(1-2\alpha -4\,\alpha^2\right. &\left.+
\cosh\left[2\,(1+\alpha)\Phi\right]\right)\nonumber
\\
&\times {\text{
sech}}^2\left[(1+\alpha)\Phi\right]\,\tanh^2\left[(1+\alpha)\Phi\right]
\label{xi2Ch}
\end{align}

%

The evolution equation for the Fourier modes of the scalar
perturbations at some scale $k$, corresponding to a comovil wave
number, is governed by the following Equation\cite{varios03} \be
\frac{d^2 u_k}{d \eta^2} + \left(k^2 -
\frac{1}{z}\frac{d^2z}{d\eta^2}\right) u_k=0, \label{uCh} \en
 where $\eta$ represents the conformal time defined by
 $\eta = \int{\frac{1}{a}\,dt}$ and $u_k$ corresponds to
 the Fourier transformed of the Mukanov variable, which is defined by
 $u = z \cal{R}$, with $z= a \frac{\dot{\phi}}{H}$ and
 $\cal{R}$ represents the gauge-invariant comovil curvature perturbation.
 This latter amount remains constant outside the
 horizon, i.e. metric perturbations with wavelengths larger than
 the Hubble radius\cite{1983}. The term $
 \frac{1}{z}\frac{d^2z}{d\eta^2}$ is usually refereed as the "mass term".

 During inflation we have that $k^2 \gg
 \frac{1}{z}\frac{d^2z}{d\eta^2}$, i.e. it is assumed that the physical
 mode has a wavelength much smaller than the curvature
 scale. Eq. (\ref{uCh}) presents a solution of the type
 \be u_k(\eta) \sim e^{-ik\eta}\left(1+\frac{{\cal
 {A}}_k}{\eta}+....\right).
\label{suCh}
 \en
Contrary, when $k^2  \ll
 \frac{1}{z}\frac{d^2z}{d\eta^2}$, the physical modes
 present wavelengths much bigger than the curvature
 scale. The term $\frac{1}{z}\frac{d^2z}{d\eta^2}$ it is found to be\cite{SL93}.
\be \frac{1}{z}\frac{d^2z}{d\eta^2} = 2 a^2 H^2\left\{ 1 +
\epsilon_{_H}
 -\frac{3}{2}\eta_{_H}- \frac{1}{2}\epsilon_{_H}
 \eta_{_H} + \frac{1}{2}\eta_{_H}^2  +
\frac{1}{2H}\dot{\epsilon}_{_H} - \frac{1}{2 H}\dot{\eta}_{_H}
 \right\}, \label{zCh}
\en

It is well known that Eq. (\ref{uCh}) solves exactly when $
\frac{1}{z}\frac{d^2z}{d\eta^2}$ is proportional to $\eta^{-2}$,
in which case Eq. (\ref{uCh}) reduces to a Bessel equation, where
the solution becomes $u_k \sim \sqrt{-k \tau} H_{\nu}(-k \tau)$,
with $H_{\nu}$ the Hankel function of first kind, and the
parameter $\nu$ depends on the slow-roll parameter $\epsilon$ via
$\nu = 3/2 + \epsilon/(1-\epsilon)$. This happens when the scale
factor expands as a power law, i.e. $a(t) \sim t^p\,\, (p > 1)$.
Here, it is obtained that $\epsilon_{_H} = \eta_{_H} =
\text{Constant}$\cite{LS92}. Others solutions, far from the
slow-roll approximation, are described in Ref. \cite{K97}. In
the generalized Chaplygin gas case this issue becomes more subtle
and needs to be worked numerically.

The power spectrum becomes defined in terms of the two point
correlation function as
\be {\cal{P}}_{\cal{R}}(k)=\frac{k^3}{2\pi^2}<
{\cal{R}}_{{\overrightarrow{k}'}} {\cal{R}}_{{\overrightarrow{k}}
}>
\delta (\overrightarrow{k}'+\overrightarrow{k} ),
\label{28}
\en
which in terms of the $u_k$ and $z$ it becomes
\be
 {\cal{P}}_{\cal{R}}(k)=\frac{k^3}{2\pi^2}\left|\frac{u_k}{z}\right|^{^{2}}.
\label{29} \en

By solving equation (\ref{uCh}) it is needed to impose boundary
conditions to the solutions. Asymptotic conditions are usually
consider to be the so-called Bunch-Davies vacuum state\cite{K12}
\be
u_k \rightarrow \left\{ \begin{array}{lll}
\frac{1}{\sqrt{2k}} e^{-i k \eta} &\hspace{0.5cm} $as$& -k \eta
\longrightarrow \infty, \\
{\cal {A}}_k z &\hspace{0.5cm} $as$  & -k \eta \longrightarrow 0.
\end{array}\right.
\label{bdCh} \en
This ensures that perturbations that are generated well inside the
horizon, i.e. in the region where $k \ll aH$, the modes approach
plane waves and those that are generated well outside the horizon,
i.e. in the region where $k \gg aH$, are fixed.

The primordial curvature perturbation are given
by\cite{Ha82,varios02}
\be {\cal{P}}_{\cal{R}}(k) =\left.
\left(\frac{H}{|\dot{\phi}|}\right)^2\,\left(\frac{H}{2\,
\pi}\right)^2\right|_{aH=k}.\label{35}
\en
This perturbation is evaluated for $aH=k$, i.e. when a given mode crosses outside
the horizon during inflation. We should notice that the modes do not evolve
outside the horizon, therefore, they kept a fixed value  after crossing the horizon
during inflation. Certainly, this value coincides with that amplitude when they cross
back inside the horizon during a later epoch.

We can introduce the scalar spectral index $n_s$ defined by
\be n_s - 1 \equiv  \frac{d \ln {\cal{P}}_{\cal{R}}}{d \ln {k}}.
\label{36}\en It is not hard to see that this quantity becomes \be n_s - 1 = 2
\eta_{_{H}} -4\,\epsilon_{_{H}}.\label{37} \en
By using expression (\ref{HCh}) we obtain that
\be n_s - 1 = \frac{3}{2}\left(3+2\,\alpha -
\cosh\left[2\,(1+\alpha)\Phi\right]\right)
\,\text{sech}^2\left[(1+\alpha)\Phi\right]. \label{nsCh} \en It is
not hard to see that $n_s$ gets the value equal to one when the
dimensionless scalar field becomes either too big, in which case
the $\text{sech}$-factor becomes vanishes, or when it gets the
value $\displaystyle \Phi_{n_s=1} =
\frac{1}{2(1+\alpha)}\,\text{arccosh} (3+2\alpha)$. The former
case is unfeasible, since the inflaton field is bounded from
above, i.e. $\Phi \leq \Phi_e$. We notice that the ratio between
$\Phi_e$ and $\Phi_{n_s=1}$ depends on the parameter $\alpha$, and
since we have considered that this parameter is located in the
range $0 < \alpha < 1$, we get that this ratio gets an upper bound
$ \left(\frac{\Phi_e}{\Phi_{n_s=1}}\right)\, <\, 1.0\,
(\text{for}\,\alpha = 1)\, \text{or}\,
\left(\frac{\Phi_e}{\Phi_{n_s=1}}\right)\,<\,1.3\,
(\text{for}\,\alpha=0)$. Thus, we see that they becomes closer
each other when the parameter $\alpha$ gets closer to one. From
this we see that this model favors a model of inflation which does
not present a Harrison-Zeld´dovic spectrum, unless the parameter
 $\alpha$ gets the value equal to one. This result is in agreement with the
 conclusion reached by {\em Planck}, which rules out exact scale invariance at over
 $5\sigma$. As we shall see, by using the data realized by {\em
 Planck} for the scalar spectral index, together with its running, will
 give a value for the generalized Chaplygin parameter $\alpha$ of the order
of $0.2$.

Let us introduce the {\em running scalar spectral index}, $n_{run}
\equiv \displaystyle \frac{d n_s}{d \ln {k}} $ which results to be
\be n_{run}=10\,\epsilon_{_{H}} \eta_{_{H}} - 8\,\epsilon_{_{H}}^2
- 2\, \xi_{_{H}}^2, \label{38} \en
that in our case it becomes
\begin{eqnarray} n_{run}= \frac{9}{4}\left(5 + 18\alpha + 16 \alpha^2 -
3\,  \cosh^2\left[2\, (1 +
   \alpha)\Phi\right]\right) & \nonumber \\
  &\times \text{sech}^2\left[(1 + \alpha)\Phi\right]
   \tanh^2\left[(1 + \alpha)\Phi\right].\label{alCh}
   \end{eqnarray} This quantity expressed in terms of the
   parameter $n_s$  becomes
\be n_{run}= \frac{1}{2}\,\frac{(4+3\,\alpha
-n_s)}{(2+\alpha)^2}\left[-10+4\,n_s + 9\,\alpha\,(1+n_s) +
8\,\alpha^2\,(2+n_s)\right]. \label{run2Ch} \en Therefore, we see
that it is possible to obtain the value of the generalized
Chaplygin gas parameter $\alpha$ if we give the corresponding
values of the parameters $n_s$ and $n_{run}$ simultaneously. Given these values,
what results is a cubic equation for the parameter $\alpha$. To do
this, we shall take the data realized by the {\em Planck}
collaboration\cite{Planck2013}, which gives the following values
for the scalar spectral index, $n_s = 0.9603 \pm 0.0073$, and the
running of the scalar spectral index, $n_{run} = - 0.0134 \pm
0.0090$. With these data, together with Eq. (\ref{run2Ch}), we
find that the possible values for the generalized Chaplygin gas
parameter are $\alpha_1 = -1.0249 \pm 0.0074$, $\alpha_2 = -0.9916
\pm 0.0048$ and $\alpha_3 = 0.2578 \pm 0.0009$. Under the
condition that this parameter should be positive we choose the
parameter $\alpha_3$\cite{W13} as the appropriate value to
consider.

Besides the scalar curvature perturbations, transverse-traceless
tensor perturbations can also be generated from quantum
fluctuations during inflation\cite{MFB92}. The tensor
perturbations do not couple to matter and consequently they are
only determined by the dynamics of the background metric. The two
independent polarizations evolve like minimally coupled massless
fields with spectrum
\begin{eqnarray}
\label{40} {\cal {P}}_{{\cal T}}= \frac{16
\pi}{m_{_{Pl}}^2}\left.\left(\frac{H}{2\pi}\right)^2
\right|_{aH=k}.
\end{eqnarray}
Equivalently to the scalar perturbations, it is possible to
introduce the {\it gravitational wave spectral index} $n_{_{T}}$
defined by $\displaystyle n_{_{T}} \equiv \frac{d \ln {{\cal
{P}}_{{\cal T}}}}{d \ln {k}}$, which in our case it becomes $
n_{_{T}} = - 2\,\epsilon_{_{H}}$. At this point we can introduce
the tensor-to-scalar amplitude ratio $\displaystyle r \equiv
\frac{{\cal {P}}_{{\cal T}}}{{\cal{P}}_{\cal{R}}}$ which becomes
\be r = 4\,\epsilon_{_{H}}. \label{rCh} \en
Thus, we obtain that
\be r= 6\,\tanh^2\left[(1+\alpha)\,\Phi\right]. \label{r2Ch} \en
Note that, since {\em Planck} has established an upper bound on
the tensor-to-scalar ratio at $r < 0.11$ (95$\%$ CL), expression
(\ref{r2Ch}) tells us that the value of the dimensionless scalar
field is bounded from above, i.e. $\Phi <
\frac{0.1326}{(1+\alpha)}$.

We may write a relationship between parameter $r$ and the
parameter $n_s$, which results to be
\be r = 6\,\left[\frac{1 + \alpha -(n_s - 1)}{2 + \alpha}\right]
\label{rns2Ch} \en

\begin{figure}[th]
\centering
\includegraphics[width=12cm,angle=0,clip=true]{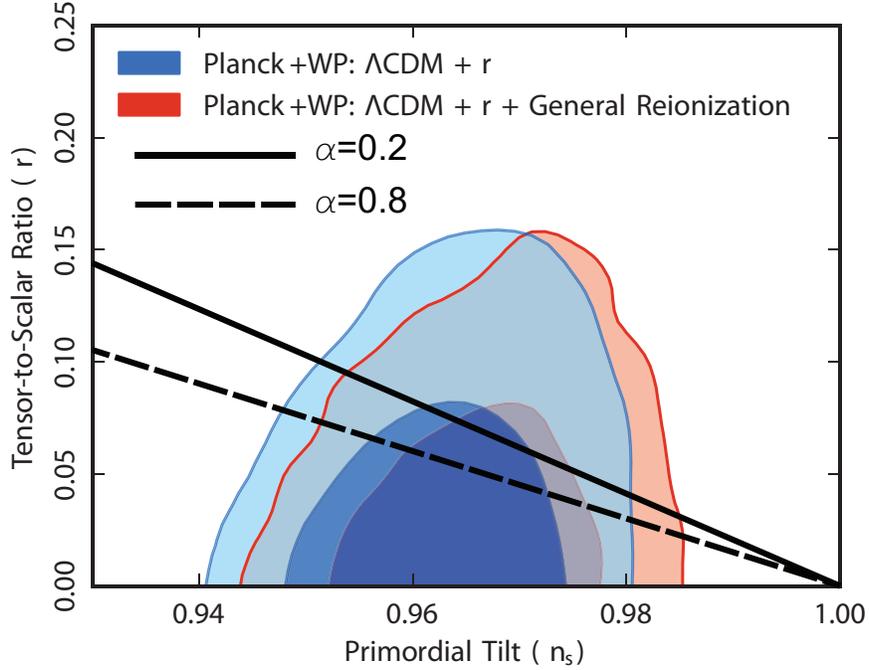}
\caption{This plot shows  the parameter $r$ as a function of the
scalar spectral index $n_s$ for two values of the constant
$\alpha$, i. e. $\alpha = 0.2$ (solid line) and $\alpha = 0.8$
(dashed line). The values of $r$ have been normalized in such a
way that we have $r=0$ when $n_s = 1$.} \label{fig3}
\end{figure}

Figure \ref{fig3} shows how changes $r$ as a function of $n_s$ for
two different values of the parameter $\alpha$. These values are
$\alpha =0.2 $ (solid line) and $\alpha =0.8$ (dashed line). Here,
we have normalized the value of the parameter $r$ in such a way
that it acquires a vanishing value when the parameter $n_s$ gets
the value one. This situation is confronted with recent data
released by {\em Planck}, where marginalized joint $68\%$ and
$95\%$ Confident Level regions for Planck plus WMAP data for the
model $\Lambda$CDM plus $r$ for instantaneous and general
reionization were considered. From this and from what we got
above, we may say that a description of inflationary universe
models in terms of a scalar inflaton field with characteristic of
a generalized Chaplygin gas could quite well accommodate  the
recently data released by the {\em Planck} mission.

Finally, combining the expression for $n_T$ together with the
expression for the $r$ parameter we get that
$r=-\frac{1}{2}\,n_T$. This expression corresponds to the
inflationary consistency condition\cite{K94}. However, this
relation could be violated in some cases\cite{H02}. Furthermore,
this consistency condition is useful to understand how $r$ is
connected to the evolution of the scalar inflaton field. It is not
hard to show that the following relation holds \be
\frac{\triangle\,\phi}{m_{Pl}} =
\frac{1}{4\,\sqrt{\pi}}\,\int_0^N\,\sqrt{r}\,dN, \label{dphi} \en
relation known as the Lyth bound\cite{L97}. As a consequence this
relation implies that an inflaton variation of the order of the
Planck mass is needed to produce $r \geq 0.01$\cite{Planck2013}.
By using the expression that we found for the $r$ parameter,
expressed by Eq. (\ref{r2Ch}), and using the result for the number
of e-folds, $N$, expressed by Eq. (\ref{N2Ch}), we get that \be
\frac{\triangle\,\phi}{m_{Pl}} =
\frac{1}{4}\,\sqrt{\frac{6}{\pi}}\,\ln\left[\frac{e^{\frac{3}{2}N}+\sqrt{e^{3N}+1}}{1+\sqrt{2}}\right].
\label{dphiCh} \en For instance, by taking some values of the
number of e-folding we get that
 $\displaystyle \frac{\triangle\,\phi}{m_{Pl}}
\approx 31$  and $\displaystyle \frac{\triangle\,\phi}{m_{Pl}}
\approx 36$, for $N=60$ and $N=70$, respectively.



\section{conclusions}

We have considered an inflationary universe model in which the
inflaton field is characterized by an Equation of state
corresponding to a generalized Chaplygin gas, i.e. $\displaystyle
p_{\phi} = - \frac{B}{\rho_{\phi}^{\alpha}}$, where $\alpha$ is
the generalized Chaplygin gas parameter, and was considered to lie
in the range $0\leq \alpha \leq 1$. In this study it was described
the kinematical evolution where the Hubble parameter was taken to
be given by $\displaystyle H(\phi) =
H_0\,\cosh^{\frac{1}{1+\alpha}}\,\left[\sqrt{\frac{6\,\pi}
{m_{Pl}^2}}\,(1+\alpha)(\phi-\phi_0)\right]$. Here, the scalar
potential, the corresponding number of e-folding and the attractor
feature of the model were described. We should mention here that
the scalar potential related to the inflaton field results in such
a way that it is possible to reproduce a generalized Chaplygin
gas, requiring that some specific initial conditions on the
inflaton field and its time derivative are chosen. Here, it was
shown that this sort of potential works quite well when this is
contrasted with the measurement recently released by the Planck
data. This situation is the main motivation to study inflationary
universe models with this kind of scalar potential.

Then, we determined the scalar and tensor spectrum indices in term
of  
$\epsilon_{_{H}}$ and $\eta_{_{H}}$ parameters. From these quantities we were
able to write down explicit expressions for the running scalar
spectral index, $n_{run}$, and the tensor-to-scalar ratio, $r$, parameters.

The resulting contours in the $r-n_s$ plane were presented for two different
values of the generalized Chaplygin gas parameter $\alpha=0.2$ and $\alpha = 0.8$.
In this plot we have confronted our results with recent data
released by {\em Planck}, where marginalized joint $68\%$ and
$95\%$ Confident Level regions for Planck plus WMAP data for the
model $\Lambda$CDM plus $r$ were used. Also, by using the values released
by {\em Planck} for the $n_s$ parameter and its running, $n_{run}$, was able to obtain
a value for the $\alpha$ parameter given by the value $\alpha = 0.2578 \pm 0.0009$.

In general terms, we have found that the tensor-to-scalar ratio can
adequately accommodate the currently available observational data
for some values of the parameter $\alpha$. In this context, it seems that
the model described here is appropriated for describing
inflationary universe models.


\begin{acknowledgments}
This work was supported by the COMISION NACIONAL DE CIENCIAS Y
TECNOLOGIA through FONDECYT Grant N$^{0}$ 1110230 and also was
partially supported by PUCV Grant N$^0$ 123.710/2011.

\end{acknowledgments}

\end{document}